\documentclass[11pt]{article}

\usepackage[margin=1.25in]{geometry}
\usepackage{newtxtext,newtxmath}
\usepackage[T1]{fontenc}
\usepackage[utf8]{inputenc}
\usepackage{lmodern}
\usepackage{setspace}
\usepackage[colorlinks=true, allcolors=black]{hyperref}
\usepackage{url}
\usepackage{booktabs}
\usepackage{longtable}
\usepackage{enumitem}
\usepackage{microtype}
\usepackage[english]{babel}
\usepackage{array}    
\usepackage{makecell} 
\newcolumntype{C}[1]{>{\centering\arraybackslash}p{#1}}   
\newcolumntype{L}[1]{>{\raggedright\arraybackslash}p{#1}}

\title{Cisco Integrated AI Security and Safety Framework Report}

\author{%
  Amy Chang\thanks{Lead author.} \quad
  Tiffany Saade\thanks{Co-author.} \quad
  Sanket Mendapara\footnotemark[2] \quad
  Adam Swanda\thanks{Contributing author.} \quad
  Ankit Garg\footnotemark[3]\\[0.5ex]
  {\normalsize Cisco AI Threat and Security Research}%
}

\date{December 2025}

\begin{document}
\maketitle

\begin{abstract}
Artificial intelligence (AI) systems are being readily and rapidly adopted, increasingly permeating critical domains---from consumer platforms and enterprise software to networked systems with embedded agents. While this has unlocked potential for human productivity gains, the attack surface has expanded accordingly: threats now span content safety failures (e.g., harmful or deceptive outputs), model and data integrity compromise (e.g., poisoning, supply-chain tampering), runtime manipulations (e.g., prompt injection, tool and agent misuse), and ecosystem risks (e.g., orchestration abuse, multi-agent collusion). Existing frameworks such as MITRE ATLAS, National Institute of Standards and Technology (NIST) AI 100-2 Adversarial Machine Learning (AML) taxonomy, and OWASP Top 10s for Large Language Models (LLMs) and Agentic AI Applications provide valuable viewpoints, but each covers only slices of this multi-dimensional space.

This paper presents Cisco’s Integrated AI Security and Safety Framework, a unified, lifecycle-aware taxonomy and operationalization framework that can be used to classify, integrate, and operationalize the full range of AI risks. It integrates AI security and AI safety across modalities, agents, pipelines, and the broader ecosystem. The AI Security Framework is designed to be practical for threat identification, red-teaming, risk prioritization, and it is comprehensive in scope and can be extensible to emerging deployments in multimodal contexts, humanoids, wearables, and sensory infrastructures. We analyze gaps in prevailing frameworks, discuss design principles for our framework, and demonstrate how the taxonomy provides structure for understanding how modern AI systems fail, how adversaries exploit these failures, and how organizations can build defenses across the AI lifecycle that evolve alongside capability advancements.
\end{abstract}

\vspace{1em}

\vspace{3em}

\begin{center}
\begin{minipage}{0.9\linewidth}
\footnotesize
{Disclaimer:} This taxonomy aims to identify certain regulatory frameworks relevant to AI safety and security objectives. Users are responsible for conducting their own assessments to ensure compliance with all applicable laws, regulations, and standards.
\end{minipage}
\end{center}

\newpage
\tableofcontents
\newpage

\section{The Evolution of AI Systems Requires Appropriate Security Paradigms}

Organizations large and small seek utility and productivity gains from AI systems, with a surveyed 69 percent of companies ranking AI as a top IT budget priority.\cite{cisco_ai_readiness} The ecosystem of AI developers and the supporting technological infrastructure (chips, data centers) are buttressed by billions of dollars of investment.\cite{wired_ai_data_centers} The speed and scale of development eclipses organizations’ abilities to shift budgets, outlooks, and training, let alone account for security risks that inevitably result from rapid AI adoption. Our own research reveals that only 33 percent of companies have a formal change management plan to guide employees through AI adoption, and only 29 percent attest that they are fully equipped against AI threats.\cite{cisco_ai_readiness} This gap exposes a fundamental limitation: humans, organizations, and governments cannot adequately comprehend or respond to the implications of such rapidly evolving technology.\cite{ubellacker2025makingsenseailimitations} While AI capabilities double in months, our cognitive frameworks, organizational processes, and regulatory mechanisms operate on timescales measured in months or years. Amidst this friction, security measures become obsolete before implementation; regulations address yesterday's threat landscape while new capabilities (and attendant threats) continue to emerge.

Understanding the challenge of trying to align security postures with the pace of AI advancement, we sought to develop a framework that measures existing risks and anticipates emerging risks in AI, serving as a bridge between AI innovation and our ability to deploy it responsibly. Securing AI systems requires a comprehensive understanding of their unique threat landscape and their contribution to the expansion of the attack surface.

Our research and development led to the creation of Cisco’s Integrated AI Security and Safety Framework (interchangeably referred to as ``AI Security Framework'' or ``Framework'' throughout this paper), a holistic framework specifically designed to address this convergence crisis across the entire AI lifecycle and across security and safety risks. The AI Security Framework is an intuitive taxonomy and operationalization strategy that integrates unique AI security threats, content and harms associated with AI inputs and outputs, and supply chain threats (Model Context Protocol (MCP), agentic, and model supply chain) into a single, comprehensive, lifecycle-aware structure that recognizes the interconnected nature of modern AI risks.\cite{anthropic_mcp}

Organizations today face a fragmented landscape of security guidance, where multiple frameworks offer partial solutions, each emphasizing distinct aspects of AI risk, such as model robustness, data privacy, or adversarial threats. Prominent frameworks such as MITRE ATLAS and OWASP Top 10 for LLMs and Generative AI catalog AI-specific threats and vulnerabilities, but face limitations in their lifecycle awareness, agentic applications, and coverage of the nexus of content harms that can elicit (or be the result of) AI-specific attacks.\cite{mitre_atlas,owasp_llm_top10,owasp_agentic_top10} Further, NIST AI 100-2’s taxonomy explicitly states that the terminology and definitions are not intended to be exhaustive.\cite{nist_ai_100_2} Frontier AI labs have similarly released guidelines and focus areas for AI security and safety, often underpinned by Responsible AI principles that they’ve drafted (e.g., Anthropic’s Constitutional AI, Google’s AI Principles).\cite{anthropic_constitutional_ai,google_saif} Table~\ref{tab:framework-comparison} below provides our assessment of the coverage areas for other AI security frameworks, and is explained in greater detail in Appendix~B.

Meanwhile, organizations continue to onboard new AI systems---from retrieval-augmented generation (RAG) systems to large language models (LLMs), AI agents, and other bespoke AI solutions---without fully understanding either unique security risks posed or the appropriate security controls to enforce, further obscuring the operating picture and complicating defensive efforts.\cite{cisco_ai_readiness} The combination of novel technology, security issues that few practitioners and researchers deeply understand, and the absence of unified approach to securing AI leaves critical vulnerabilities unaddressed.

\begin{table}[htbp]
  \centering
  \small 
  \caption{Assessment of coverage across AI security taxonomies and frameworks}
  \label{tab:framework-comparison}
  \begin{tabular}{L{3.5cm} C{1.5cm} C{1.5cm} C{1.5cm} C{1.5cm} C{2.8cm}}
    \toprule
    \makecell[b]{Dimension} &
    \makecell{MITRE\\ATLAS} &
    \makecell{NIST\\AML} &
    OWASP &
    Industry &
    \makecell{Cisco AI\\Security\\Framework} \\
    \midrule
    Content safety        & No     & Partial & No     & Partial & Yes \\
    AI security           & Yes    & Partial & Yes    & Partial & Yes \\
    Lifecycle scope       & Partial& Yes     & Partial& Partial & Yes \\
    Multi-agent/tools     & No     & Partial & Partial& Partial & Yes \\
    Multi-modal           & No     & Partial & No     & Partial & Yes \\
    Supply chain          & Partial& Partial & Partial& Partial & Yes \\
    Unified integration   & No     & Partial & No     & No      & Yes \\
    \bottomrule
  \end{tabular}
\end{table}

The AI Security Framework is our effort to bridge those gaps. It was designed to incorporate five essential dimensions that reflect the complexity of contemporary AI systems, including: the integration of AI threats and content harms, AI development lifecycle awareness, multi-agent coordination, multi-modality, and audience-aware utility. The following sections dive deeper into our framework methodology and discuss how our Taxonomy operationalizes defense-in-depth principles across AI development, deployment, and post-deployment operations. This report then describes immediate operationalization opportunities for the Framework, and opportunities for continued extensibility and application. We hope this brings together AI leaders, developers, security practitioners, policymakers, and enthusiasts alike to identify additional areas of development and work together towards a safer, more secure AI future.

\section{Cisco Integrated AI Security and Safety Framework}

\subsection{What are AI Security and AI Safety?}

With AI systems operating at increasing scales and levels of autonomy, the need for precise conceptual frameworks around aspects of AI risk has grown. Yet the terms ``AI security'' and ``AI safety'' remain inconsistently defined across researchers, policymakers, and practitioners. The concepts evolve alongside the technology itself, shaped by shifting threats, novel capabilities, and the realization that AI introduces challenges not fully captured by traditional security or risk frameworks.

It is with this foundation that we define AI security as:

\medskip

\emph{the discipline of ensuring AI accountability and protecting AI systems from unauthorized use, availability attacks, and integrity compromise across the AI lifecycle.}

\medskip

Our definition draws from traditional cybersecurity principles of confidentiality, integrity, and availability of AI systems, but we emphasize the importance of accountability as a distinct and essential component of AI security. As agentic frameworks and autonomous tooling proliferate (that is, systems that can listen, reason, and act autonomously to achieve goals with limited human oversight), humans must maintain accountability across these systems. Extending human permissions, accesses, and resources to these systems requires mechanisms that ensure their actions remain transparent, auditable, and controllable.

Contemporary discussions of AI safety began to emerge around mid-2010s, as rapid advancements in AI/ML prompted questions about the technology’s impact on society.\cite{amodei2016concreteproblemsaisafety, russell_dewey_tegmark} With the rise of generative AI and increasingly capable models, discourse on AI safety continued to expand, as the risks and potential consequences have increased in tandem.\cite{bengio2025internationalscientificreportsafety} We define AI safety as:

\medskip

\emph{helping ensure AI systems behave ethically, reliably, fairly, transparently, and in alignment with human values.}

\medskip

Taken together, AI security and AI safety form complementary dimensions of a unified risk framework: one concerned with protecting AI systems from threats, and the other with ensuring that their behavior remains aligned with human values and ethics. Treating these domains in tandem can enable organizations to build AI systems that are not only robust and reliable, but also responsible and worthy of trust.

\subsection{Design Principles}

The Integrated AI Security and Safety Framework is built upon five design elements that distinguish it from prior taxonomic efforts and attempt to encompass an evolving AI threat landscape: integration of AI threats and content harms, AI development lifecycle awareness, multi-agent coordination, multi-modality, and audience-aware utility.

\paragraph{Integration of AI threats and content harms.} Our framework accounts for harmful and toxic content alongside attacks against AI systems and components. Rather than treating these as separate concerns (as many existing frameworks do), the AI Security Framework recognizes that adversaries exploit vulnerabilities across both domains, and oftentimes, link content manipulation with technical exploits to achieve their objectives. For example, an attacker might use prompt injection to bypass content filters while simultaneously exfiltrating proprietary training data, which demonstrates how both content and security threats converge in real-world attack scenarios.

\paragraph{Lifecycle awareness.} Rather than focusing solely on deployment or runtime concerns, the AI Security Framework provides visibility across the entire AI lifecycle: from data collection and curation to training and validation, deployment and integration, runtime operations, and incident response and remediation. Just like threats are manifested and implemented differently at each stage, defenses that are effective at one phase may be irrelevant or insufficient at another. Therefore, by explicitly covering each lifecycle phase, the framework allows customers to implement defense-in-depth strategies that account for how risks evolve as AI systems progress from development to production and beyond. For example, data poisoning attacks occur during training, while prompt injection threatens deployed systems, which reveals how context fundamentally shapes both threat vectors and appropriate mitigations that customers would need to implement.

\paragraph{Multi-agentic and tool-aware.} The AI Security Framework accounts for the risks that emerge when AI systems work together, encompassing orchestration patterns, inter-agent communication protocols, shared memory architectures, and collaborative decision-making processes.\cite{raza2025trismagenticaireview} Our taxonomy accounts for associated risks that emerge in systems with autonomous planning capabilities (agents), external tool access (MCP), persistent memory, and multi-agent collaboration---threats that would be invisible to or underrepresented in frameworks designed for earlier generations of AI technology.

\paragraph{Multi-modality.} The framework addresses threats across all data types and interaction modes beyond just text-based models, to include code, images, audio, video, sensor streams, and embedded control systems. This facilitates comprehensive coverage as AI systems increasingly operate across multiple modalities simultaneously. This breadth supports applicability to diverse AI deployments and use-cases/products, ranging from conversational assistants to autonomous vehicles. More specifically, malicious objectives and techniques that affect image classifiers differ fundamentally from prompt injections in language models, yet both represent adversarial input manipulation requiring systematic categorization.

\paragraph{An AI security compass.} The AI Security Framework serves as both an operational framework and a conceptual map: it provides teams with a shared language and mental model for understanding the threat landscape beyond individual model architectures. The framework includes the supporting infrastructure, complex supply chains, organizational policies, and human-in-the-loop interactions that collectively determine security outcomes. This can enable clearer communication between AI developers, AI end-users, business functions, security practitioners, and governance and compliance entities. A suitable example of this is the taxonomy’s coverage of supply chain attacks targeting model repositories, compromised training infrastructure, or backdoored dependencies, which represent ecosystem-level threats that individual model security cannot address.

The Framework is also an operational tool: offering practical guidance for implementing technical controls, designing comprehensive monitoring systems, structuring incident response playbooks, and allocating resources effectively across the risk landscape. The framework recognizes that human operators, feedback mechanisms, and organizational governance structures introduce both vulnerabilities and opportunities for defense, requiring security strategies that account for the full range of technical and human elements in AI deployment contexts.

\subsection{Taxonomy Structure}

Our taxonomy is comprised of two components designed to help organizations systematically identify, prioritize, and mitigate risk across the AI lifecycle. The first component covers technical threats to AI systems, including LLMs, agentic systems, and elements of AI supply chain (from data pipelines to model artifacts to deployment environments). The second addresses harmful and toxic content that users might input or that AI systems might generate. Because LLMs and other generative systems can both process and produce text, images, videos, and audio, our safety framework enumerates 25 harmful content and safety categories.

As the taxonomy encompasses generative, agentic, and supply chain threats, it provides organizations with an actionable foundation for proactively managing risk at scale. Teams can use the taxonomy to:
\begin{itemize}[leftmargin=*]
  \item Assess which objectives, techniques, and subtechniques map to their risk appetite, industry requirements, and business priorities;
  \item Design targeted red-team evaluations and security tests;
  \item Identify the most relevant content-safety categories for monitoring and filtering; and
  \item Benchmark system maturity and guide investment toward the highest-priority risks.
\end{itemize}

Our taxonomy will be dynamically updated as novel threats emerge or previously theoretical threats materialize. The AI Security Framework will track and reflect the changing threat landscape, and we intend to publish best practices, identify known indicators associated with specific threats, and surface mitigation techniques across core domains, including:
\begin{itemize}[leftmargin=*]
  \item Content safety (e.g., output filtering, policy enforcement);
  \item Runtime security (e.g., prompt injection detection, jailbreak prevention);
  \item Agentic security (e.g., tool mediation, memory hygiene, agent sandboxing);
  \item Data security (e.g., RAG integrity, vector DB protection, data leakage prevention);
  \item Supply chain security (e.g., model scanning, dependency analysis, provenance tracking); and
  \item Infrastructure security (e.g., compute quotas, secret management, network segmentation).
\end{itemize}

With the AI Security Framework, we aim to equip organizations to prioritize the threats that matter most and implement robust, end-to-end defenses for modern AI systems.

\subsection{AI Security Framework Taxonomy Hierarchy}

Our AI Security Framework’s taxonomy operates on four hierarchical levels (objectives, techniques, subtechniques, and procedures), each providing progressively granular detail about adversarial threats. For ease of navigation and comprehension, we’ve grouped objectives into three risk groups: common manipulation threats, data-related threats, and downstream threats and impact risks. This structure can enable organizations to understand both the strategic intent behind attacks and the tactical methods used to execute them, facilitating comprehensive threat modeling and targeted defense strategies. As of our initial release, our taxonomy is comprised of:
\begin{itemize}[leftmargin=*]
  \item 19 Objectives: High-level attacker goals;
  \item 40 Techniques: Specific attack methods;
  \item 112 Subtechniques: Detailed implementation patterns; and
  \item Procedures: Specific and discrete instructions and tools used to conduct attack techniques and achieve objectives.\footnote{Procedures are not counted because specific implementation instructions are already quite numerous and rapidly changing, such that it would not be a worthwhile endeavor to comprehensively catalog.}
\end{itemize}

\subsubsection*{Objectives: The ``Why'' Behind Attacks}

Objectives represent the ultimate goals or motives driving adversarial activity. They represent what an attacker aims to accomplish through their actions. We have identified 19 distinct objectives that encompass the range of malicious goals observed across the AI threat landscape. These objectives serve as the foundation for threat modeling, allowing organizations to prioritize defenses based on which adversarial goals pose the greatest risk to their specific systems and operations. Each objective is designated with an OB-XXX identifier (e.g., OB-001 through OB-019) and represents a distinct category of adversarial intent.

\paragraph{Example: OB-004: Communication Compromise.}

``Communication Compromise'' represents attacks where the primary goal is to intercept, alter or impersonate communications between an AI system and its users. This objective encompasses various attack scenarios including:
\begin{itemize}[leftmargin=*]
  \item Man-in-the-middle attacks on agent communications;
  \item Exploit unclear or porous boundaries between different system components using message injection or modification;
  \item Malicious impersonation of users, agents, or system components; and
  \item Tampering with tool calls or API responses.
\end{itemize}

Organizations concerned about protecting the integrity and confidentiality of their AI system communications would prioritize testing against OB-004, which includes all associated techniques and subtechniques designed to achieve this objective. As organizations scope testing scenarios against OB-004, organizations may further consider specific contexts such as:
\begin{itemize}[leftmargin=*]
  \item Business priorities: what operations or data are most critical to protect?
  \item Risk appetite: what level of risk is acceptable for different or overlapping system functions?
  \item Industry sector: what are the most prevalent threats in my domain (e.g., finance, healthcare, entertainment)?
  \item Regulatory or governance considerations: what compliance or governance obligations exist and must be met?
  \item System architecture: how do I map my AI deployment attack surface?
\end{itemize}

These then influence considerations for AI red teaming, AI threat modeling, runtime monitoring and guardrail deployment, controls implementation, and resource allocation. As the threat landscape evolves, these objectives can be periodically reviewed to maintain comprehensive coverage, and testing outcomes can help organizations identify strengths and areas for improvement in their AI defenses, enabling data-driven improvements to security posture.

\subsubsection*{Techniques: The ``How'' of Attack Execution}

Techniques represent the specific methods, processes, or actions that adversaries employ to achieve their objectives. While objectives describe what an attacker wants to accomplish, techniques describe how they accomplish it. Our taxonomy currently includes 40 distinct techniques, each mapped to an objective. Each technique is designated with a hierarchical identifier (e.g., AITech-1.1, AITech-1.2, AITech-2.1, AITech-2.2) where the first number corresponds to the primary objective and the second number indicates the specific technique within that objective category.

\paragraph{Example: AITech-1.1: Direct Prompt Injection.}

``Direct Prompt Injection'' is one of four primary techniques that can be used to achieve the Goal Hijacking objective (OB-001). This technique involves directly manipulating user-provided prompts or instructions to override system directives, bypass safety constraints, or redirect the model's behavior toward attacker-defined objectives.

While our taxonomy shows a one-to-one mapping between each technique and its associated objective, it is important to understand that techniques often support multiple objectives or serve as intermediate steps in complex attack chains. In practice, certain techniques such as ``AITech-1.3: Multi-Agent Prompt Injection,'' ``AITech-7.3: Token Source Manipulation,'' ``AITech-9.2: Detection Evasion'' are mapped to the following objectives, respectively: ``OB-001: Goal Hijacking,'' ``OB-007: Sabotage/Integrity Loss,'' and ``OB-009: Supply Chain Compromise,'' these can support multiple other objectives or serve as one of several steps adversaries use to achieve their objective. To illustrate this further:

\begin{itemize}[leftmargin=*]
  \item AITech-1.3: Multi-Agent Prompt Injection: Primarily mapped to Goal Hijacking (OB-001), but can also enable:
    \begin{itemize}[leftmargin=*]
      \item Data theft (OB-002) by manipulating agents to exfiltrate information;
      \item System manipulation (OB-005) by coordinating malicious agent behaviors; and
      \item Privilege escalation (OB-014) by compromising high-privilege agents.
    \end{itemize}
  \item AITech-9.2: Detection Evasion: Primarily mapped to Supply Chain Compromise (OB-009), but serves as an enabling technique for:
    \begin{itemize}[leftmargin=*]
      \item Nearly all other objectives by helping attacks avoid detection;
      \item Persistent access scenarios; and
      \item Long-term compromise operations.
    \end{itemize}
\end{itemize}

When determining which techniques to test against, organizations should leverage the AI Security Framework to map techniques to priority objectives, focusing on those that are most relevant to that organization’s risk profile and technology infrastructure. Organizations can then leverage threat intelligence to identify recent attack patterns, industry-specific threats, and emerging techniques observed in the wild. After, they can assess existing controls to identify which techniques your current defenses are least prepared to handle (e.g., techniques with no existing mitigations, techniques that bypass current controls, and techniques that combine multiple attack vectors) against your specific AI architecture (e.g., multi-agent systems, RAG systems, tool-using agents, and API-based AI systems). This focused approach enables testing efforts to concentrate on the most probable and highest-impact adversarial techniques for their specific context.

\subsubsection*{Subtechniques: Granular Attack Variants}

Subtechniques are the more detailed level of the taxonomy, capturing specific variations, implementations, or specialized methods within each technique. These represent the diverse ways a technique can be executed in practice, accounting for different attack vectors, technical approaches, and contextual variations. Our taxonomy currently includes 112 subtechniques that offer precise visibility into adversarial AI attacks. Each subtechnique is designated with a three-level hierarchical identifier (e.g., AISubtech-3.1.1, AISubtech-3.1.2) where: the first number is the objective category, the second number is the specific technique identifier, and the third number is the subtechnique variant.

Breaking techniques down to the subtechnique level can provide several critical advantages:
\begin{itemize}[leftmargin=*]
  \item Facilitate precise threat identification to pinpoint exactly which attack variations organizations are susceptible to;
  \item Tailor security controls and defensive measures to specific attack methods, improving detection accuracy and controls coverage;
  \item Guide comprehensive red-teaming or evaluation coverage to cover numerous variations of a technique;
  \item Catalyze threat research through tracking specific attack patterns, understand evolving adversary methodologies, or share more precise threat indicators; and
  \item Simplify deeper risk assessments that can enable more accurate prioritization and resource allocation.
\end{itemize}

\paragraph{Example: Impersonation Technique Breakdown.}

Consider AITech-3.1 Masquerading / Obfuscation / Impersonation under the objective (OB-003) with the same name. This technique encompasses various methods adversaries use to assume false identities within AI systems. The taxonomy breaks this down into specialized subtechniques:

\medskip

\noindent\textbf{AISubtech-3.1.1: Identity Obfuscation}

\emph{Definition:} Manipulating agent or user identity representation within context, metadata, or interaction patterns to evade detection, tracking, or access controls.

\emph{Attack methods:}
\begin{itemize}[leftmargin=*]
  \item False identity claims in prompts or context;
  \item Spoofed metadata and authentication headers;
  \item Identity rotation to avoid pattern detection;
  \item Anonymous/pseudonymous operation to prevent attribution; and
  \item Credential obfuscation techniques.
\end{itemize}

\emph{Example scenario:} An attacker manipulates agent metadata to claim administrator identity, bypassing role-based access controls to access restricted tools and data.

\medskip

\noindent\textbf{AISubtech-3.1.2: Trusted Agent Spoofing}

\emph{Definition:} Masquerading as a legitimate agent or MCP-registered service to inject malicious instructions, responses, or outputs that other agents, services, or users treat as trusted. This attack exploits the assumption of authenticity within multi-agent systems and protocol-mediated toolchains, enabling actors to bypass safeguards and influence downstream behavior.

\emph{Attack methods:}
\begin{itemize}[leftmargin=*]
  \item Name collision with legitimate agents;
  \item Spoofed agent credentials or cryptographic signatures;
  \item Fake tool registration in MCP servers;
  \item Man-in-the-middle positioning as legitimate agent; and
  \item Response injection from ``trusted'' sources.
\end{itemize}

\emph{Example scenario:} An attacker registers a malicious tool with the same name as a legitimate database query tool but at higher priority. When agents call ``database\_query,'' they connect to the attacker’s tool instead, enabling data exfiltration and query manipulation.

By breaking it down to the subtechnique level, the taxonomy and broader Framework provide a more nuanced explanation of the diverse ways impersonation attacks can manifest, helping support more targeted detection, prevention, and response strategies.

\subsubsection*{Procedures: Precise Implementation Details}

Procedures represent the most granular level of the taxonomy, capturing the exact step-by-step implementations, specific tools, code patterns, and precise technical methods used to execute subtechniques in practice. Drawn from the traditional cybersecurity tactics, techniques, and procedures (TTP) lens, procedures answer the question, ``How exactly is this done?'' They document the specific technical details that distinguish one implementation from another, even when executing the same subtechnique.

Objectives, techniques, and subtechniques describe classes of adversarial behavior, and procedures document instances of actual implementation. Examples of procedures include:
\begin{itemize}[leftmargin=*]
  \item Specific encoding methods: Base64, ROT13, hexadecimal, Unicode escaping;
  \item Particular evasion patterns: character substitution schemes, linguistic tricks, formatting variations;
  \item Exact tool usage: specific scripts, frameworks, or platforms employed;
  \item Code-level details: actual payload structures, API calls, parameter combinations; and
  \item Environmental prerequisites: required conditions, configurations, or dependencies.
\end{itemize}

Due to the variability and near-constant evolving nature of procedure-level attacks, they are not explicitly captured in the taxonomy. This level is inherently dynamic, requiring continuous updates; we call out explicit examples across our taxonomy in the form of indicators and examples for specific subtechniques and modalities (e.g., MCP, agentic) that we’ve observed in-the-wild or as a result of our security research and red teaming efforts.

\subsection{Model Context Protocol Threats Taxonomy}

Model Context Protocol threats represent a critical layer of lifecycle-aware AI security: the protocol governs how LLMs interpret tools, prompts, metadata, and execution environments. When MCP components are tampered with, impersonated, or misused, benign agent operations can be redirected toward data exposure, unauthorized system access, code execution, or malicious tool invocation. MCP threats arise when untrusted inputs, deceptive decorators, manipulated tool definitions, or tampered execution environments cause the model to perform actions outside its intended scope.

Our MCP threats taxonomy currently covers 14 threat types and maps to corresponding AI Security Framework objectives, techniques, and subtechniques, with relevant indicators, severity levels, and mitigation best practices, and fall into four major groups:
\begin{itemize}[leftmargin=*]
  \item Injection and Interpretation Threats (3 threat types) that mislead the model into executing harmful or unintended tool calls.
  \item Tool Integrity Threats (3 threat types) where malicious or spoofed tools replace trusted logic, intercept data, or embed hidden behavior.
  \item Data Exfiltration and Access Threats (4 threat types) enabling unauthorized file-system access, internal service reachability, arbitrary resource read/write privilege manipulation, or leakage of secrets and proprietary data.
  \item Execution and Payload Threats (4 threat types) enable arbitrary or hidden execution of code within the MCP environment, stemming from unsafe evaluation constructs, insecure tool definitions, dynamic imports, or serialized payload loading.
\end{itemize}

\subsection{Supply Chain Threats Taxonomy}

Supply chain risk is a core dimension of lifecycle-aware AI security: tampered artifacts, dependencies, or build pipelines can convert benign models into vectors for code execution, data exfiltration, or harmful outputs. Our AI supply chain threat taxonomy is designed to be simultaneously scan-friendly and governance-ready: every threat maps back to the AI Security Framework objectives, techniques, and subtechniques, and points to the most relevant file types, indicators, severity, and an actionable ``Model Defense Layer'' mitigation approach. Our supply chain threat taxonomy is organized into four primary groups encompassing 22 distinct threat types:

\begin{itemize}[leftmargin=*]
  \item Artifact and Format Vulnerabilities (7 threat types);
  \item Model Manipulation and Tampering (5 threat types);
  \item Dependency and Distribution Compromise (5 threat types); and
  \item Operational and Runtime Threats (5 threat types).
\end{itemize}

\paragraph{Artifact and Format Vulnerabilities.} Vulnerabilities that are intentionally embedded directly in model files, tokenizers, archives, or metadata that permit code execution or hidden triggers. The types of threats include: deserialization or serialized-code execution; format-specific backdoors (e.g., Lambda layers); tokenizer or template injection (e.g., server-side template injection); hidden native libraries or executables; archive or compression abuses; metadata/manifest tampering; and format inconsistencies and poisoned arrays.

\paragraph{Model Manipulation and Tampering.} These threats encompass intentional changes to training data, weights, or adapters that create bias, trojans, or stealthy triggers. The types of threats include: training data poisoning; weight poisoning or backdoor insertion; malicious adapters (e.g., low-rank adaptation or parameter-efficient fine-tuning) or merged-model trojans; annotation or label tampering; and snapshot or shard manipulation.

\paragraph{Dependency and Distribution Compromise.} These threats arise from packages, registries, mirrors, and continuous integration/continuous deployment (CI/CD) scenarios that produce or distribute tainted artifacts. Threats in this category include: malicious package/tool injection; registry/mirror compromise (rug pulls); typosquatting/namespace squatting; dependency replacement/version downgrade attacks; and CI/CD and artifact-swap compromises.

\paragraph{Operational and Runtime Threats.} These threats include runtime behaviors that turn supply chain compromises into operational impact. Threats in this category include: arbitrary code execution on load; unauthorized system access or privilege escalation; unauthorized network access or data exfiltration; model extraction or weight reconstruction; runtime obfuscation or evasion; and telemetry or forensics tampering.

\subsection{AI Safety: Harmful Content Taxonomy}

In addition to the adversarial threat taxonomy focused on attack techniques and security vulnerabilities, organizations must also address the fundamental challenge of harmful content generation. These are outputs that, while not necessarily the result of adversarial attacks, can cause significant harm to users, organizations, and society. Our AI Safety: Harmful Content taxonomy provides a structured way to identify dangerous or undesirable AI-generated outputs across LLMs and multimodal systems. The AI security components of the Framework also intersect: an adversary might use prompt injection techniques to force a model to generate hate speech, for instance, combining a security attack with a safety violation. Understanding both taxonomies is essential for comprehensive AI governance.

Our team conducted an extensive review of research, open-source datasets (both academic and from industry), published standards, and existing guardrail solutions to understand and unify the content moderation taxonomies into a single aggregate taxonomy that balances across content severity, viability of detection, hierarchical classification, and real-world examples to derive useful information. The below content categories were the result of our research, and represent the 25 categories (sorted into five primary groups) with the most egregious harm and with the categories with the most unique content distributions:

\begin{itemize}[leftmargin=*]
  \item Cybersecurity and Hacking (2 types);
  \item Safety Harms and Toxicity (16 types);
  \item Integrity Compromise (4 types);
  \item Intellectual Property Compromise (2 types); and
  \item Privacy Attacks (1 type).
\end{itemize}

These categories help organizations establish clear evaluation criteria for generative outputs, enforce safety policies, and design effective guardrail and content moderation strategies that work across text, image, audio, video, and other modalities.

\subsubsection*{Cybersecurity and Hacking}

This category captures content that facilitates cyber attacks, system compromise, or malicious technical activities. This group contains two primary content types that represent distinct but related threats to digital security: malware or exploits (AI-generated or AI-assisted code or software aimed at facilitating cyberattacks) and cyber abuse (AI-enabled manipulation to gain unauthorized access or cause harm such as social engineering), both defined in greater detail below, and both having observed implementations over the past year.\cite{anthropic_ai_espionage, google_threat_ai} Due to the dual-use nature of cybersecurity knowledge, the line between education and enablement can be subtle. Guardrails or other defensive measures here should carefully calibrate their safety policies to allow legitimate security discourse while preventing the generation of weaponized offensive content.

\paragraph{Malware and Exploits.} Malware and exploits represent one of the most technically sophisticated categories of harmful content. This encompasses the generation of malicious code designed to compromise computer systems, including viruses, ransomware, trojans, and worms. It could even extend to the creation of exploits targeting known or zero-day vulnerabilities, detailed instructions for developing attack tools, and obfuscated or polymorphic malware variants designed to evade detection.

\paragraph{Cyber Abuse.} Cyber abuse captures a broader range of malicious technical activities beyond malware development, and could include detailed technical guidance for unauthorized system access, network penetration methodologies, credential theft techniques, authentication bypass methods, social engineering attacks methods, distributed denial-of-service attack coordination.

\subsubsection*{Safety Harms and Toxicity}

The largest category, capturing sixteen content categories that pose direct physical, psychological, or societal harm to individuals or groups, and encompasses:
\begin{itemize}[leftmargin=*]
  \item Animal Abuse;
  \item Child Abuse / Exploitation;
  \item Disinformation;
  \item Environmental Harm;
  \item Financial Harm;
  \item Harassment;
  \item Hate Speech;
  \item Non-Violent Crime;
  \item Profanity;
  \item Scams and Deception;
  \item Self-Harm;
  \item Sexual Content and Exploitation;
  \item Social Division and Polarization;
  \item Terrorism / Extremism;
  \item Violence and Public Safety Threat; and
  \item Weapons / Chemical, Biological, Radiological, and Nuclear (CBRN) Risks.
\end{itemize}

\subsubsection*{Integrity Compromise}

This category captures content that undermines trust, provides unauthorized professional guidance, or spreads falsehoods. Unlike the ``Safety Harms'' group above, which focuses on direct physical or psychological harm, the Integrity Compromise group captures harms related to misinformation, professional boundaries, and institutional trust, and includes categories such as:
\begin{itemize}[leftmargin=*]
  \item Hallucinations / Misinformation;
  \item Unauthorized Financial Advice;
  \item Unauthorized Legal Advice; and
  \item Unauthorized Medical Advice.
\end{itemize}

\subsubsection*{Intellectual Property Compromise}

This category captures content that attempts to enable, promote, or facilitate unauthorized use, reproduction, or distribution of copyrighted or trademarked material, as well as any data that is not meant for public release or disclosure (e.g., business processes and plans, contracts, internal manuals, credentials).

\subsubsection*{Privacy Attacks}

This final category addresses content that violates data privacy. This group represents increasingly critical concerns as AI systems handle more data and demonstrate greater capability to extract, infer, or expose private information, such as personally identifiable information (PII), protected health information (PHI), or payment card industry (PCI) information.

\subsubsection*{Concerns and Considerations for Multi-Modal Complexity}

While the categories themselves remain consistent across modalities, the technical challenges of detection and enforcement vary significantly by content type. Text-based hate speech might be detected through natural language processing or ML classification, but the same hateful content embedded in an image requires computer vision capabilities to extract text or recognize symbols. We assess this will increasingly become challenging as generative AI capabilities continue to advance, but advancements in interpretability research, including techniques that identify and manipulate how concepts are represented internally, may yield modality-agnostic approaches to detection and enforcement.

By organizing harmful content into five groups encompassing 25 distinct content types—from cybersecurity threats and safety harms to integrity violations and privacy attacks—the taxonomy enables clear evaluation criteria, comprehensive policy development, and effective content moderation strategies across all modalities.

\section{Operationalizing the AI Security Framework}

The Integrated AI Security and Safety Framework provides organizations with a flexible, scalable basis for AI security and safety that is operationalizable across strategic, operational, and tactical levels of AI security operations. Each objective within the taxonomy can be mapped to specific preventive, detective, and corrective controls, facilitating risk management and integration with organizational policies and governance structures. While our initial release does not comprehensively capture these, we intend to continue to build out these elements of our taxonomy. Meanwhile, objectives and techniques can serve as the foundation for test case templates, enabling red-teaming exercises, and providing clear coverage indicators.

Furthermore, the Framework provides common risk language, correlated detections, unified incident response, and comprehensive coverage mapping across multiple vectors to support prioritization, detection, and mitigation efforts, such as:
\begin{itemize}[leftmargin=*]
  \item Content safety: implementing output filtering and policy enforcement;
  \item Runtime security: creating policies or guardrails to detect against prompt injections and jailbreaks;
  \item Agentic security: crafting tool and behavior profiles to control agentic tool use, support agent and memory hygiene, or compartmentalizing agent actions within secured or segmented areas;
  \item Data security: confirming RAG integrity, protecting vector databases, and preventing unwanted data leakage;
  \item Supply chain security: analyzing and tracking model integrity, auditing dependencies and model provenance for tampering; and
  \item Infrastructure security: enforcing compute quotas, secrets management, and network segmentation configurations.
\end{itemize}

Let’s expand on an operationalization example that is at the intersection of AI security and AI safety taxonomies. We understand that adversaries often use threat techniques specifically to force generation of safety-violating content: a jailbreak attack (AITech-2.1) may be employed specifically to make a system generate hate speech, malware code, or disinformation; a direct prompt injection attack (AITech-1.1) could be used to override safety controls and generate child abuse content or terrorism materials. Understanding these intersections helps organizations build defenses that address both the attack vectors and the harmful outcomes. Adversarial safety testing could (and should) specifically target the combination of threat techniques and safety categories: for instance, testing whether various jailbreak methods can force generation of each content type in the safety taxonomy.

We’ve also built out dedicated MCP and model supply chain taxonomies that seamlessly map with the broader Framework, facilitating awareness of threats and vulnerabilities across the AI lifecycle. (We intend to supplement these taxonomies shortly following the publication of this report with an agentic taxonomy.) The AI supply chain taxonomy, for example, details threats to AI model files, and includes vulnerabilities such as deserialization risks, architectural backdoors, data poisoning, and supply chain compromises in model artifacts, flagging for risks that arise from the use of first- and third-party models, outdated dependencies, and adversarial manipulation that could lead to biased outputs, system compromise, or data exfiltration. In practice, the taxonomy assists in the identification, detection, assessment, and mitigation of these threats.

The Framework’s use of common risk language also helps facilitate incident reporting and information sharing for AI security: allowing consistent and efficient descriptions of incidents and threat information to be shared both within and across organizations.\cite{wei2025designingincidentreportingsystems} The adaptability of this framework can help technical and business stakeholders communicate more effectively about AI security.

\subsection{Connection to Global Policy Frameworks and AI Regulations}

The introduction and adoption of AI regulations and laws at both in the United States and internationally have the potential to reshape how organizations measure and secure themselves against AI risk. Our taxonomy is designed not only to comprehensively capture the spectrum of threats, but also to help organizations build more robust risk management programs for AI security amidst a growing landscape of AI legal and regulatory requirements.

\subsubsection*{Connection to Regulatory Frameworks}

Consider how the Goal Hijacking objective (OB-001) relates to concerns highlighted in multiple international instruments. The Council of Europe Framework Convention on Artificial Intelligence and Human Rights, Democracy and the Rule of Law and the Budapest Convention on Cybercrime both address unauthorized manipulation of computer systems.\cite{coe_ai_convention,budapest_convention} The Budapest Convention requires countries that ratify the Convention to criminalize illegal access (Article 2), data interference (Article 4), and system interference (Article 5), and specifically to address unauthorized access into computers or the alteration, deletion or poisoning of data or system functionalities.

Organizations that operate in countries where the Budapest Convention has been ratified could use our taxonomy to help understand how the Convention may impact their AI systems. For example, Goal Hijacking represents an AI-specific example of illegal access, data interference, and system interference. Our taxonomy further captures the unique nature of AI attacks, where adversaries subvert intelligent systems by corrupting their goals rather than their code. The Budapest Convention’s prohibition against ``system interference'' (Article 5) and ``misuse of devices'' (Article 6) can also be applied to an AI context, where an attacker coerces misalignment of a model, impairing its normal system functioning. The Framework applies this concept while further identifying AI-specific attack vectors, such as subverting an AI system's intended function for malicious ends.

\subsubsection*{Broader Regulatory Relevance and Standards Alignment}

The AI Security Framework is also designed to give organizations concrete examples of how AI-specific threats and attacker objectives may pertain to best practices, standards, and regulations, both AI-specific and more broadly. Some examples of how our taxonomies relate to different regulatory contexts and standards include:

\begin{itemize}[leftmargin=*]
  \item \textbf{Risk Management Frameworks:} NIST AI Risk Management Framework (AI RMF) aligns objectives and techniques with the Govern, Map, Measure, and Manage functions, providing concrete threat scenarios for each risk category for AI risk management.\cite{nist_ai_rmf}
  \item \textbf{Regulations:} The European Union (EU) AI Act identifies areas relevant to AI security risk (e.g., Articles 9 (Risk Management System), 15 (Accuracy, Robustness, and Cybersecurity), and 17 (Quality Management System)).\cite{eu_ai_act} For Article 9, the Framework can provide a framework to assist in identifying, analyzing, and documenting AI-specific risks as part of a risk management system. For Article 15, the Framework may help organizations identify specific security requirements that may be considered in ensuring an accurate, robust, and secure AI system, such as protecting against System Manipulation (OB-005) and Supply Chain Compromise (OB-009).
  \item \textbf{Laws and Directives:} The EU’s updated Network and Information Systems Directive (NIS2) similarly covers risk management and incident reporting requirements for what the NIS2 describes as ``important'' and ``essential'' entities.\cite{nis2} The Framework can support structured categorization to classify incidents by objective, technique, and subtechnique, aiding organizations in understanding regulatory requirements and regulatory compliance systems.
\end{itemize}

The Framework can help with risk assessment and documentation, controls mapping, incident response and reporting, and cross-jurisdictional coherence for multi-national organizations. As AI regulation evolves, the AI Security Framework can serve as a bridge between existing security practices and emerging legal requirements. It grounds adversarial behaviors in a structured framework that relates to concepts in established regulatory conventions (i.e., cybersecurity and traditional security frameworks) while capturing AI-specific threats.

\subsection{Our Vision for Standardization and Integration}

Realizing the full potential of the Integrated AI Security and Safety Framework requires comprehensive standardization efforts that bridge the gap between taxonomic classification and practical implementation. These efforts span numerous interconnected domains, each essential to creating a mature, interoperable AI security ecosystem. While we present the blueprint for how we envision the standardization and integration of our taxonomy below, it will require our broader technology, security, governance and policy communities to come together to work towards the common goal of securing AI.

\subsubsection*{Control Catalogs and Regulatory Principles}

Control catalogs represent the operational foundation that translates categories within a taxonomy into concrete protective measures. Following the structural principles of the NIST Cybersecurity Framework\cite{nist_cyberframework} and NIST 800-171\cite{nist_800_171_r2} security guidelines, these catalogs map specific security controls to each objective, technique, and subtechnique in the taxonomy. This mapping can help organizations to systematically select appropriate controls based on their threat profile, align with regulatory principles, and translate high-level threat awareness into actionable defensive measures. By standardizing control descriptions and implementation guidance, the catalogs support consistency across deployments while accommodating organization-specific customization.

\subsubsection*{Standardized Evaluation and Testing}

Building upon these control foundations, standardized evaluation suites provide the means to verify security effectiveness. These suites include comprehensive benchmarks and adversarial test harnesses covering techniques and subtechniques across the AI Security Framework, supported by automated evaluation pipelines that enable continuous security assessment. By standardizing what to test (shaped by the taxonomy) and how to test (evaluation methodologies), organizations transform security assessment from ad-hoc penetration testing into rigorous, repeatable processes. This standardization enables meaningful comparison of security postures across different AI systems, providers, and deployment contexts---critical for procurement decisions, third-party risk assessment, and demonstrating due diligence to regulators and stakeholders.

\subsubsection*{Provenance and Supply Chain Integrity}

As AI systems are more deeply intertwined in our technology stacks, provenance standards for establishing trust and traceability are critical. These standards encompass cryptographic signing, hardware-based attestation, and extensions to traditional Software Bill of Materials frameworks. Nascent attempts at capturing AI Bills of Materials (AIBOMs) focus on provenance information including model origins, training data lineage, and complete dependency chains. This transparency enables security teams to identify supply chain risks and verify the integrity of AI components throughout their lifecycle.

\subsubsection*{Security Ecosystem Interoperability}

Standardization can enable ecosystem-wide interoperability where security controls, evaluation results, and provenance information can be shared and verified across organizational boundaries. In cybersecurity, one of the most mature and widely adopted approaches to structured and unified information exchange is the combination of Structured Threat Information eXpression (STIX) and Trusted Automated eXchange of Intelligence Information (TAXII).\cite{oasis_stix, oasis_taxii} STIX provides a standardized, machine-readable language that uses a common syntax to describe cyber threat information and mitigation steps. TAXII complements this by defining the protocols and services for securely transporting that intelligence between organizations. By extending STIX/TAXII with fields and ontologies aligned to the AI Security Framework, organizations can also represent AI-specific threats, such as model poisoning attempts and prompt injection patterns, in a format that existing cybersecurity information sharing infrastructure already understands.

This approach bolsters collective defense against AI threats by enabling dissemination of emerging threats or effective countermeasures across organizations and sectors.

\section{AI Advancement Will Continue: So Will Our Framework}

As AI systems grow more complex and adversarial techniques evolve, we intend to keep up. While the Integrated AI Security and Safety Framework represents the most comprehensive AI threat taxonomy to date, a rapidly evolving AI landscape demands continuous extension, updates, and operationalization. New attack techniques or content types may emerge as AI capabilities expand into new domains. Existing threat and content categories may require refinement as our understanding of AI security deepens through real-world experience. The boundaries between categories may shift, and the appropriate severity levels and responses may need adjustment based on changing technologies, regulatory requirements, and capabilities. Organizations should view this taxonomy not as a static checklist but as a living framework that provides structure while remaining adaptable to new challenges. The work of building safe AI systems is ongoing, but frameworks like this taxonomy provide essential structure for that critical endeavor.

We alone cannot solve AI security. The AI Security Framework provides depth in multi-agent coordination, tool augmentation, multi-modal attacks, and agentic supply chain threats, but many challenges in the AI threat landscape remain, and additional threats have yet to surface. We hope that with the introduction of Cisco’s Integrated AI Security and Safety Framework, we educate AI security professionals, provide baseline best practice guidelines, and inspire other AI security practitioners to contribute and build upon this taxonomy.

\newpage
\appendix

\section{Appendix A: Integrated AI Security and Safety Framework Taxonomy}

You may find the most up-to-date AI security and safety taxonomy here:\cite{cisco_taxonomy}
\begin{center}
\url{https://www.cisco.com/site/us/en/learn/topics/artificial-intelligence/ai-safety-security-taxonomy.html}
\end{center}

\noindent The following summarizes the taxonomy identifiers:

\medskip
\noindent OB-001: Goal Hijacking

\noindent\quad AITech-1.1: Direct Prompt Injection

\noindent\qquad AISubtech-1.1.1: Instruction Manipulation (Direct Prompt Injection)

\noindent\qquad AISubtech-1.1.2: Obfuscation (Direct Prompt Injection)

\noindent\qquad AISubtech-1.1.3: Multi-Agent Prompt Injection

\noindent\quad AITech-1.2: Indirect Prompt Injection

\noindent\qquad AISubtech-1.2.1: Instruction Manipulation (Indirect Prompt Injection)

\noindent\qquad AISubtech-1.2.2: Obfuscation (Indirect Prompt Injection)

\noindent\qquad AISubtech-1.2.3: Multi-Agent (Indirect Prompt Injection)

\noindent\quad AITech-1.3: Goal Manipulation

\noindent\qquad AISubtech-1.3.1: Goal Manipulation (Models, Agents)

\noindent\qquad AISubtech-1.3.2: Goal Manipulation (Tools, Prompts, Resources)

\noindent\quad AITech-1.4: Multi-Modal Injection and Manipulation

\noindent\qquad AISubtech-1.4.1: Image-Text Injection

\noindent\qquad AISubtech-1.4.2: Image Manipulation

\noindent\qquad AISubtech-1.4.3: Audio Command Injection

\noindent\qquad AISubtech-1.4.4: Video Overlay Manipulation

\medskip
\noindent OB-002: Jailbreak

\noindent\quad AITech-2.1: Jailbreak

\noindent\qquad AISubtech-2.1.1: Context Manipulation (Jailbreak)

\noindent\qquad AISubtech-2.1.2: Obfuscation (Jailbreak)

\noindent\qquad AISubtech-2.1.3: Semantic Manipulation (Jailbreak)

\noindent\qquad AISubtech-2.1.4: Token Exploitation (Jailbreak)

\noindent\qquad AISubtech-2.1.5: Multi-Agent Jailbreak Collaboration

\medskip
\noindent OB-003: Masquerading / Obfuscation / Impersonation

\noindent\quad AITech-3.1: Masquerading / Obfuscation / Impersonation

\noindent\qquad AISubtech-3.1.1: Identity Obfuscation

\noindent\qquad AISubtech-3.1.2: Trusted Agent Spoofing

\medskip
\noindent OB-004: Communication Compromise

\noindent\quad AITech-4.1: Agent Injection

\noindent\qquad AISubtech-4.1.1: Rogue Agent Introduction

\noindent\quad AITech-4.2: Context Boundary Attacks

\noindent\qquad AISubtech-4.2.1: Context Window Exploitation

\noindent\qquad AISubtech-4.2.2: Session Boundary Violation

\noindent\quad AITech-4.3: Protocol Manipulation

\noindent\qquad AISubtech-4.3.1: Schema Inconsistencies

\noindent\qquad AISubtech-4.3.2: Namespace Collision

\noindent\qquad AISubtech-4.3.3: Server Rebinding Attack

\noindent\qquad AISubtech-4.3.4: Replay Exploitation

\noindent\qquad AISubtech-4.3.5: Capability Inflation

\noindent\qquad AISubtech-4.3.6: Cross-Origin Exploitation

\medskip
\noindent OB-005: Persistence

\noindent\quad AITech-5.1: Memory System Persistence

\noindent\qquad AISubtech-5.1.1: Long-term / Short-term Memory Injection

\noindent\quad AITech-5.2: Configuration Persistence

\noindent\qquad AISubtech-5.2.1: Agent Profile Tampering

\medskip
\noindent OB-006: Feedback Loop Manipulation

\noindent\quad AITech-6.1: Training Data Poisoning

\noindent\qquad AISubtech-6.1.1: Knowledge Base Poisoning

\noindent\qquad AISubtech-6.1.2: Reinforcement Biasing

\noindent\qquad AISubtech-6.1.3: Reinforcement Signal Corruption

\medskip
\noindent OB-007: Sabotage / Integrity Degradation

\noindent\quad AITech-7.1: Reasoning Corruption

\noindent\quad AITech-7.2: Memory System Corruption

\noindent\qquad AISubtech-7.2.1: Memory Anchor Attacks

\noindent\qquad AISubtech-7.2.2: Memory Index Manipulation

\noindent\quad AITech-7.3: Data Source Abuse and Manipulation

\noindent\qquad AISubtech-7.3.1: Corrupted Third-Party Data

\noindent\quad AITech-7.4: Token Manipulation

\noindent\qquad AISubtech-7.4.1: Token Theft

\medskip
\noindent OB-008: Data Privacy Violations

\noindent\quad AITech-8.1: Membership Inference

\noindent\qquad AISubtech-8.1.1: Presence Detection

\noindent\quad AITech-8.2: Data Exfiltration / Exposure

\noindent\qquad AISubtech-8.2.1: Training Data Exposure

\noindent\qquad AISubtech-8.2.2: LLM Data Leakage

\noindent\qquad AISubtech-8.2.3: Data Exfiltration via Agent Tooling

\noindent\quad AITech-8.3: Information Disclosure

\noindent\qquad AISubtech-8.3.1: Tool Metadata Exposure

\noindent\qquad AISubtech-8.3.2: System Information Leakage

\noindent\quad AITech-8.4: Prompt/Meta Extraction

\noindent\qquad AISubtech-8.4.1: System LLM Prompt Leakage

\medskip
\noindent OB-009: Supply Chain Compromise

\noindent\quad AITech-9.1: Model or Agentic System Manipulation

\noindent\qquad AISubtech-9.1.1: Code Execution

\noindent\qquad AISubtech-9.1.2: Unauthorized or Unsolicited System Access

\noindent\qquad AISubtech-9.1.3: Unauthorized or Unsolicited Network Access

\noindent\qquad AISubtech-9.1.4: Injection Attacks (e.g., SQL, Command Execution, XSS)

\noindent\qquad AISubtech-9.1.5: Template Injection (e.g., SSTI)

\noindent\quad AITech-9.2: Detection Evasion

\noindent\qquad AISubtech-9.2.1: Obfuscation Vulnerabilities

\noindent\qquad AISubtech-9.2.2: Backdoors and Trojans

\noindent\quad AITech-9.3: Dependency / Plugin Compromise

\noindent\qquad AISubtech-9.3.1: Malicious Package / Tool Injection

\noindent\qquad AISubtech-9.3.2: Dependency Name Squatting (Tools / Servers)

\noindent\qquad AISubtech-9.3.3: Dependency Replacement / Rug Pull

\medskip
\noindent OB-010: Model Theft / Extraction

\noindent\quad AITech-10.1: Model Extraction

\noindent\qquad AISubtech-10.1.1: API Query Stealing

\noindent\qquad AISubtech-10.1.2: Weight Reconstruction

\noindent\qquad AISubtech-10.1.3: Sensitive Data Reconstruction

\noindent\quad AITech-10.2.1: Model Inversion

\medskip
\noindent OB-011: Adversarial Evasion

\noindent\quad AITech-11.1: Environment-Aware Evasion

\noindent\qquad AISubtech-11.1.1: Agent-Specific Evasion

\noindent\qquad AISubtech-11.1.2: Tool-Scoped Evasion

\noindent\qquad AISubtech-11.1.3: Environment-Scoped Payloads

\noindent\qquad AISubtech-11.1.4: Defense-Aware Payloads

\noindent\quad AITech-11.2: Model-Selective Evasion

\noindent\qquad AISubtech-11.2.1: Targeted Model Fingerprinting

\noindent\qquad AISubtech-11.2.2: Conditional Attack Execution

\medskip
\noindent OB-012: Action-Space and Integration Abuse

\noindent\quad AITech-12.1: Tool Exploitation

\noindent\qquad AISubtech-12.1.1: Parameter Manipulation

\noindent\qquad AISubtech-12.1.2: Tool Poisoning

\noindent\qquad AISubtech-12.1.3: Unsafe System / Browser / File Execution

\noindent\qquad AISubtech-12.1.4: Tool Shadowing

\noindent\quad AITech-12.2: Insecure Output Handling

\noindent\qquad AISubtech-12.2.1: Code Detection / Malicious Code Output

\medskip
\noindent OB-013: Availability Abuse

\noindent\quad AITech-13.1: Disruption of Availability

\noindent\qquad AISubtech-13.1.1: Compute Exhaustion

\noindent\qquad AISubtech-13.1.2: Memory Flooding

\noindent\qquad AISubtech-13.1.3: Model Denial of Service

\noindent\qquad AISubtech-13.1.4: Application Denial of Service

\noindent\qquad AISubtech-13.1.5: Decision Paralysis Attacks

\noindent\quad AITech-13.2: Cost Harvesting / Repurposing

\noindent\qquad AISubtech-13.2.1: Service Misuse for Cost Inflation

\medskip
\noindent OB-014: Privilege Compromise

\noindent\quad AITech-14.1: Unauthorized Access

\noindent\qquad AISubtech-14.1.1: Credential Theft

\noindent\qquad AISubtech-14.1.2: Insufficient Access Controls

\noindent\quad AITech-14.2: Abuse of Delegated Authority

\noindent\qquad AISubtech-14.2.1: Permission Escalation via Delegation

\medskip
\noindent OB-015: Harmful / Misleading / Inaccurate Content

\noindent\quad AITech-15.1: Harmful Content

\noindent\qquad AISubtech-15.1.1: Cybersecurity and Hacking: Malware / Exploits

\noindent\qquad AISubtech-15.1.2: Cybersecurity and Hacking: Cyber Abuse

\noindent\qquad AISubtech-15.1.3: Safety Harms and Toxicity: Animal Abuse

\noindent\qquad AISubtech-15.1.4: Safety Harms and Toxicity: Child Abuse / Exploitation

\noindent\qquad AISubtech-15.1.5: Safety Harms and Toxicity: Disinformation

\noindent\qquad AISubtech-15.1.6: Safety Harms and Toxicity: Environmental Harm

\noindent\qquad AISubtech-15.1.7: Safety Harms and Toxicity: Financial Harm

\noindent\qquad AISubtech-15.1.8: Safety Harms and Toxicity: Harassment

\noindent\qquad AISubtech-15.1.9: Safety Harms and Toxicity: Hate Speech

\noindent\qquad AISubtech-15.1.10: Safety Harms and Toxicity: Non-Violent Crime

\noindent\qquad AISubtech-15.1.11: Safety Harms and Toxicity: Profanity

\noindent\qquad AISubtech-15.1.12: Safety Harms and Toxicity: Scams and Deception

\noindent\qquad AISubtech-15.1.13: Safety Harms and Toxicity: Self Harm

\noindent\qquad AISubtech-15.1.14: Safety Harms and Toxicity: Sexual Content and Exploitation

\noindent\qquad AISubtech-15.1.15: Safety Harms and Toxicity: Social Division and Polarization

\noindent\qquad AISubtech-15.1.16: Safety Harms and Toxicity: Terrorism / Extremism

\noindent\qquad AISubtech-15.1.17: Safety Harms and Toxicity: Violence and Public Safety Threat

\noindent\qquad AISubtech-15.1.18: Safety Harms and Toxicity: Weapons / CBRN Risks

\noindent\qquad AISubtech-15.1.19: Integrity: Hallucinations / Misinformation

\noindent\qquad AISubtech-15.1.20: Integrity: Unauthorized Financial Advice

\noindent\qquad AISubtech-15.1.21: Integrity: Unauthorized Legal Advice

\noindent\qquad AISubtech-15.1.22: Integrity: Unauthorized Medical Advice

\noindent\qquad AISubtech-15.1.23: Intellectual Property Compromise: Intellectual Property Infringement

\noindent\qquad AISubtech-15.1.24: Intellectual Property Compromise: Confidential Data

\noindent\qquad AISubtech-15.1.25: Privacy Attacks: PII / PHI / PCI

\medskip
\noindent OB-016: Surveillance

\noindent\quad AITech-16.1: Eavesdropping

\noindent\qquad AISubtech-16.1.1: Logging Sensitive Conversations

\medskip
\noindent OB-017: Cyber-Physical / Sensor Attacks

\noindent\quad AITech-17.1: Sensor Spoofing

\noindent\qquad AISubtech-17.1.1: Sensor Spoofing: Action Signals (audio, visual)

\medskip
\noindent OB-018: System Misuse / Malicious Application

\noindent\quad AITech-18.1: Fraudulent Use

\noindent\qquad AISubtech-18.1.1: Spam / Scam / Social Engineering Generation

\noindent\quad AITech-18.2: Malicious Workflows

\noindent\qquad AISubtech-18.2.1: Abuse of APIs for Mass Automation

\noindent\qquad AISubtech-18.2.2: Dedicated Malicious Server or Infrastructure

\medskip
\noindent OB-019: Multi-Modal / Cross-Modal Risks

\noindent\quad AITech-19.1: Cross-Modal Inconsistency Exploits

\noindent\qquad AISubtech-19.1.1: Contradictory Inputs Attack

\noindent\qquad AISubtech-19.1.2: Modality Skewing

\noindent\quad AITech-19.2: Fusion Payload Split

\noindent\qquad AISubtech-19.2.1: Convergence Payload Injection

\noindent\qquad AISubtech-19.2.2: Chained Payload Execution

\newpage
\section{Appendix B: Analysis of Existing Security and Technical Standards and Frameworks}

This Appendix provides extended analysis of industry-leading and widely adopted standards in AI safety and security, as well as frontier lab approaches to AI safety and security.

\subsection*{MITRE}

MITRE is one of the leading organizations to have released frameworks to understand and assess the techniques, tactics, and procedures behind a cyber-attack on any system (MITRE ATT\&CK), but also against AI systems specifically (ATLAS).\cite{mitre_atlas} ATLAS represents one of the most comprehensive efforts to catalog adversarial tactics and techniques specific to data, models, and AI pipelines. Built on the ATT\&CK framework's methodology, ATLAS organizes threats into a matrix structure with tactics and techniques that support red-teaming, threat modeling, and security evaluation. However, ATLAS faces some limitations in their incomplete coverage of rapidly evolving AI threats, resulting in gaps in coverage of content safety concerns, multi-modal attacks, and agentic system orchestration. ATLAS also relies on documented attacks rather than forward-looking scenarios. ATLAS does not prescribe controls or governance practices, meaning organizations must combine it with broader Responsible AI or security frameworks to achieve comprehensive protection.

\subsection*{NIST AI Risk Management Framework and Adversarial Machine Learning}

The NIST AI Risk Management Framework (AI RMF) and Adversarial Machine Learning (AML) taxonomy together provide a broad foundation for understanding and mitigating risks in AI systems.\cite{nist_ai_rmf,nist_ai_100_2} The AI RMF is a voluntary, risk-based framework organized around four core functions across the AI lifecycle: Map, Measure, Manage, and Govern. The NIST RMF lifecycle governance-centric focus addresses key pillars such as validity, safety, security, privacy, and fairness. However, this framework-oriented approach, while fundamental for organizational strategy, does not provide full operational guidance and lacks the granular, threat-enumerative detail security practitioners need for day-to-day threat modeling and response, especially in a rapidly evolving threat landscape.

Complementing this, the NIST AML taxonomy offers more technical depth with coverage of adversarial attack types and phases (including evasion, poisoning, privacy attacks, and model theft) with consistent technical definitions and outlined mitigation strategies. While the NIST AML's analytical foundation supports integration into risk management programs, it remains primarily technical and threat-focused, with less emphasis on operationalization, governance, and lifecycle management. Combined, these NIST frameworks provide strong conceptual grounding, but fall short of keeping up with the evolving AI threat landscape and latest domain-specific operational frameworks to provide complete threat coverage.

\subsection*{OWASP Top 10 for LLM and GenAI Applications and Top 10 for Agentic Applications}

OWASP has also spearheaded two AI security taxonomic efforts: Top 10 for LLM and GenAI Applications and OWASP Top 10 for Agentic Applications.\cite{owasp_llm_top10,owasp_agentic_top10}  The OWASP Top 10 for LLM/GenAI represents a high-level enumeration of critical security risks such as prompt injection, insecure output handling, supply chain vulnerabilities, and sensitive information disclosure.

The Top 10 for Agentic Applications extends upon generative AI threats by identifying key security risks specific to agentic and multi-agent AI systems such as agent goal hijacking, tool misuse and exploitation, memory and context poisoning. This Top 10 for Agentic sheds light on the unique risks arising from autonomous decision-making, delegation, and tool integration.

Both lists translate diffuse research findings into accessible, testable categories, helping security teams design controls, perform threat modeling, and run red-teaming exercises. While the Top 10 serves as a practical guide for AI developers and security practitioners, the scope of the framework is intentionally (and narrowly) limited to top ten most prominent risks, with minimal lifecycle coverage and limited content safety coverage, inhibiting a comprehensive view of existing and emerging AI threats. It also lacks granularity on attack manifestation and is difficult to operationalize from a mitigating controls perspective.

\subsection*{Industry Guidelines}

Several U.S.-based frontier AI laboratories have published security guidelines, red-teaming methodologies, and responsible AI principles informed by their operational experiences. We provide a high-level overview of some prominent AI lab approaches to AI safety and security:

\begin{itemize}[leftmargin=*]
  \item OpenAI employs iterative deployment with external red-teaming to identify jailbreaks, prompt injections, and misuse vectors before broad release. Their Model Spec documentation defines how OpenAI’s AI models should behave, including the desired core behavioral principles along safety and alignment objectives. OpenAI also publishes system cards alongside model releases that document (among other things) risks and safety concerns, results of red-teaming evaluations, and mitigations implemented.\cite{openai_modelspec}
  \item Anthropic embeds safety constraints during training through Constitutional AI, its core safety and alignment principles, to guide model behavior and reduce harmful outputs. Their Responsible Scaling Policy also adjusts safety and security protections to prevent misuse as model capabilities increase, with a focus on preventing catastrophic misuse.\cite{anthropic_constitutional_ai}
  \item Microsoft's Responsible AI Standard guides AI development across six core principles: fairness; reliability and safety; privacy and security; inclusiveness; transparency; and accountability. Microsoft also conducts threat modeling and security impact assessments for AI deployments: their Counterfit tooling project provides adversarial ML testing for threats and vulnerability scanning for AI systems.\cite{microsoft_responsible_ai_standard}
  \item Google's core AI principles are designed to ensure that AI is innovative, responsible, and collectively beneficial. These principles also inform their security frameworks like its Secure AI Framework (SAIF) that details Google’s security-oriented approach to AI risk management. SAIF addresses AI-specific risks including data poisoning, prompt injection, model exfiltration or tampering, and insecure model output.\cite{google_saif}
\end{itemize}

Industry-developed AI safety and security guidelines offer several notable strengths. Because they emerge from organizations operating AI systems at scale, they incorporate practical insights that academic research alone cannot fully capture. These frameworks typically integrate security considerations with responsible AI principles, underscoring that technical robustness and ethical alignment reinforce one another. Frontier labs also introduce tooling like adversarial testing frameworks, fairness evaluation libraries, and structured red-teaming methodologies that lower the barrier to adoption and help standardize defensive practices across the ecosystem. The red-teaming approaches, in particular, have been refined through multiple deployment cycles, giving practitioners actionable models for adversarial evaluation.

These efforts do have identifiable limitations. Guidelines originating from labs are often tailored to specific products, architectures, or deployment environments which may not translate cleanly into broader, ecosystem-wide standards. Much of the underlying methodology remains proprietary or disclosed only at a high level, which constrains reproducibility and limits independent verification. The absence of a unified taxonomy across organizations also makes it difficult to compare threat categories or assess relative security posture in a consistent way. Their coverage of emerging risks—such as multi-agent systems and agentic supply chains—remains limited, as well.

\bigskip
\newpage
\nocite{*}
\bibliographystyle{plain}
\bibliography{main}

\end{document}